\documentstyle[epsf]{mn}
\begin{document}

\title[Flux-dependent amplitude of broadband noise in XRBs and AGN]
{The flux-dependent amplitude of broadband noise variability in X-ray 
binaries and active galaxies}
\author[Philip Uttley and Ian M. M$^{\rm c}$Hardy]
{Philip Uttley\thanks{e-mail: pu@astro.soton.ac.uk} and
Ian M. M$^{\rm c}$Hardy \\
Department of Physics and Astronomy, University of Southampton, 
Southampton SO17 1BJ \\
}

\date{}

\maketitle
\parindent 18pt
\begin{abstract} 
Standard shot-noise models, which seek to explain the broadband 
noise variability that characterises the X-ray lightcurves of X-ray
binaries and active galaxies, predict that the power spectrum of the
X-ray lightcurve is stationary (i.e. constant amplitude and shape) on
short time-scales.  We show that the broadband noise power spectra of
the black hole candidate Cyg~X-1 and the accreting
millisecond pulsar SAX~J1808.4-3658 are intrinsically non-stationary, in
that RMS variability scales linearly with flux. 
Flux-selected power spectra confirm that this effect is
due to changes in power-spectral amplitude and not shape.  The
lightcurves of three Seyfert galaxies are also consistent with a
linear relationship between RMS variability and flux, suggesting that it
is an intrinsic feature of the broadband noise variability in compact
accreting systems over more than 6 decades of central object mass.
The RMS variability responds to flux variations on all measured
time-scales, raising fundamental difficulties for shot-noise models
which seek to explain this result by invoking variations in the shot
parameters. 
We suggest that models should be explored where the
longest time-scale variations are fundamental and
precede the variations on shorter time-scales. 
Possible models which can explain the linear RMS-flux relation include
the fractal break-up of
large coronal flares, or the propagation of fluctuations in mass accretion rate
through the accretion disk. 
The linear relationship between RMS variability and
flux in Cyg~X-1 and SAX~J1808.4-3658 is offset on the
flux axis, suggesting the presence of a second, constant-flux component
to the lightcurve which contributes $\sim25$\% of the total flux.  The
spectrum of this constant component
is similar to the total spectrum, suggesting that it may correspond to
quiet, non-varying regions in the X-ray emitting corona.
\end{abstract}


\section{Introduction}
The broadband noise variability which characterises the X-ray lightcurves of
both X-ray binary systems (XRBs) and active galactic nuclei (AGN) is
commonly modelled as a shot-noise process.  The simplest shot-noise
models, where the lightcurve is constructed from a stochastic series of 
independent overlapping shots with a single characteristic decay time-scale 
(e.g. Terrell 1972), produce a red-noise type power spectrum at 
high frequencies ($P(\nu)\propto\nu^{-\alpha}$, where $\alpha=2$), which
flattens to $\alpha=0$ at low frequencies (i.e. the variability becomes
white-noise on long time-scales).  This simple form is 
qualitatively similar to the power-spectral shapes of XRBs, which
flatten below $\sim0.1$~Hz
(e.g. van der Klis 1995) and AGN, which flatten on much longer
time-scales ($<10^{-6}$~Hz, M$^{\rm c}$Hardy 1988, Edelson \& Nandra 1999),
but fails to reproduce the typically observed red-noise
slopes, $\alpha\sim1.0$.  More complex shot-noise models, which
invoke a broad distribution of decay time-scales, can reproduce the
correct high-frequency shapes (Lehto 1989; Lochner, Swank \& Szymkowiak
1991), including the additional high-frequency
break which is seen above $\sim1$~Hz in some XRB power
spectra (e.g. Cyg~X-1, Belloni \&
Hasinger 1990).  Physical interpretations of shots typically invoke
magnetic flares in the X-ray emitting corona which is likely to
be common to both XRBs and AGN (e.g. Poutanen \& Fabian 1999).  \\
It is well-known that the power spectra of XRBs are non-stationary (i.e.
have a variable shape and amplitude) on long time-scales
($>$days, e.g. Belloni \& Hasinger 1990).  There is some evidence for
non-stationarity of AGN power spectra on much longer time-scales (Uttley
et al. 1999).  The shot-noise models described above can
account for non-stationary power-spectra by invoking changes in the
parameters of the shot-noise process on long time-scales, but predict that 
on short time-scales the power spectrum should be stationary.  Because
individual power-spectra measured on very short time-scales (seconds) are noisy,
this prediction is difficult to test.  However, we can test whether 
power-spectral shape and/or amplitude varies systematically with short 
time-scale variations in X-ray flux by binning the power-spectra of small segments of
observed XRB and AGN lightcurves according to their flux. \\
The relationship between long-time-scale variations in flux and
power-spectral amplitude has been studied in a few XRBs, e.g.
a somewhat complex correlation between RMS variability and flux
has been noted during the decay of the hard X-ray
transient GRO~J0422+32 (Denis et al. 1994).
Surprisingly, the possibility of systematic short-time-scale variations
of power-spectral shape and amplitude with
flux has not previously been investigated in detail.  The standard practice of
normalising power spectra by squared mean flux (and RMS variability by
mean flux), for comparison between different objects and instruments,
causes information regarding the flux to be lost.  In this letter, we
show that this practice has helped to keep hidden 
a startling fact about the X-ray
variability of XRBs and AGN: that the RMS variability
of their broadband noise component scales with flux variations on
time-scales as short as seconds, while
the power-spectral shape remains constant.  This discovery poses
particular problems for standard shot-noise models, as we shall discuss.
\begin{figure*}
\begin{center}
{\epsfxsize 0.9\hsize
 \leavevmode
 \epsffile{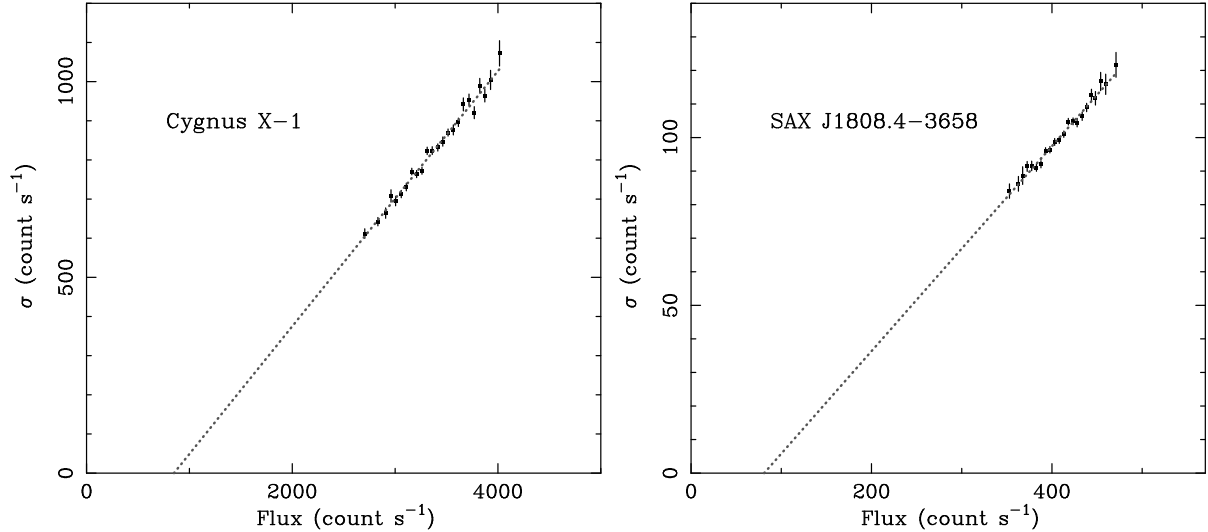}
}\caption{Local flux dependence of mean $\sigma$ for Cyg~X-1 and
SAX~J1808.4-3658.  The dotted lines mark the best-fitting 
linear models described in the text.} \label{fig:sig}
\end{center}
\end{figure*}
\section{The flux-dependent variability of X-ray Binaries}
We obtained public archival data from Rossi X-ray Timing Explorer ({\it RXTE})
observations of Cyg~X-1
(in the low state),
observed 23 October 1996 (Nowak et al. 1999) and the recently discovered
accreting millisecond pulsar, SAX~J1808.4-3658 (Wijnands \& van der Klis
1998a), observed 18 April 1998, of duration 18~ksec and 23~ksec
respectively (using the PCA instrument and including only
periods where all 5 Proportional Counter
Units (PCUs) were switched on). 
Cyg~X-1 is an obvious choice for study, as it is the
most well-known black hole X-ray binary.  The accreting millisecond
pulsar SAX~J1808.4-3658 is a good example of a neutron star system which
displays strong broadband noise variability,
so any similarity in the variations of the power spectra of these
objects will provide strong evidence that the power-spectral variability
is intrinsic to the process that produces the broadband noise,
independent of whether the central object has a hard surface or an event
horizon.  Furthermore, the
observations we select show no evidence for discernable low-frequency 
quasi-periodic oscillations
(QPOs) or components other than the broadband noise which we wish to
investigate here. 
Using PCA binned and event mode data from the observations of Cyg~X-1
and SAX~J1808.4-3658 respectively, we made lightcurves with a resolution
of 16~ms for
each source in the 2--13.1~keV energy range.
The mean fluxes in this energy range are
3377~count~s$^{-1}$ and 409~count~s$^{-1}$ for Cyg~X-1 and
SAX~J1808.4-3658 respectively. \\
We first investigated the dependence of RMS variability on local
X-ray flux, in other words, does the RMS variability of a small section of
lightcurve depend on the mean flux of that section of lightcurve?
In order to answer this question, we split each lightcurve into 10~s
segments and measured the power spectrum for each segment (subtracting the
contribution due to Poisson noise estimated from the mean total count
rate of the lightcurve segment).  We applied the standard RMS-squared
normalisation to the power spectrum, so that integrating the power over
a given frequency range yields the contribution to lightcurve variance
due to variations in that frequency range.  Taking the square root of
the integrated power yields the RMS variability $\sigma$.  By using this
method, rather than simply measuring $\sigma$ directly from the
lightcurve, we can study how the amplitude of the power spectrum in a
defined frequency range responds to changes in flux. 
We first chose to measure $\sigma$ over the frequency range 0.1-10~Hz, which incorporates
the full spectrum of broadband noise from white-noise to red-noise, as
can be seen from the power spectra of these observations (Nowak et al.
1999, Wijnands \& van der Klis 1998b).
Due to the stochastic
nature of red-noise lightcurves, there is significant scatter in
individual measures of $\sigma$, so we binned $\sigma$
as a function of segment flux using relatively
narrow flux bins to examine the form of the $\sigma$-flux
relationship over a broad range of flux.  Using a minimum number of
30 measurements per bin to ensure an accurate estimate
of the standard error in the mean $\sigma$, the resulting $\sigma$-flux
relation for both sources is plotted in Fig.~\ref{fig:sig}. 
The lightcurves of both Cyg~X-1 and SAX~J1808.4-3658 show a remarkably
linear dependence of RMS variability on flux. \\
Visual inspection of Fig.~\ref{fig:sig} shows that the $\sigma$-flux trend in
both sources does not pass through the origin, i.e. there is a constant
offset in flux and possibly also $\sigma$.  To test the goodness of fit
of a linear model, including a constant offset from the origin, we fitted
a function of the form $\sigma =k(F-C)$, where $F$ is the flux and $k$
and $C$ are constants.  This linear model provides a good fit to the
data, yielding $\chi^{2}$ values of 28.7 (for 22 degrees of freedom) for
Cyg~X-1 and 17.2 (20 degrees of freedom) for SAX~J1808.4-3658.  The
best fitting model parameters were $k=0.326\pm0.017$,
$C=850\pm130$~count~s$^{-1}$ for Cyg~X-1 and $k=0.305\pm0.026$,
$C=81\pm30$~count~s$^{-1}$ for SAX~J1808.4-3658 (uncertainties are at
the 90\% confidence level for 2 interesting parameters).  Note that the
gradient of the $\sigma$-$F$ trend $k$, is equivalent to the fractional
RMS variability of the
variable-$\sigma$ component of the lightcurve.  The constant $C$
represents a second component to the lightcurve, which does not follow
the linear $\sigma$-$F$ trend but may still contribute to the total
value of $\sigma$.  It is not possible to disentangle the mean flux
level of this second, constant-$\sigma$ component and the contribution
it makes to the total $\sigma$, however the value of $C$ represents the
flux of this component in the limit where it does not vary. \\
\begin{figure}
\begin{center}
{\epsfxsize 0.9\hsize
 \leavevmode
 \epsffile{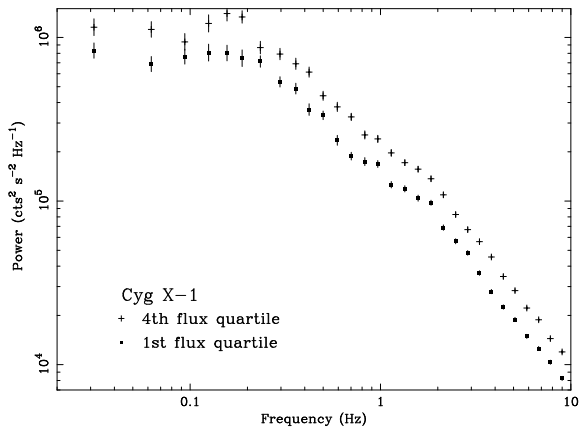}
}\caption{Flux-dependent power spectra of Cyg~X-1, corresponding 
to the first and fourth
flux quartiles (see text for details).} \label{fig:fluxps}
\end{center}
\end{figure}
Changes in RMS variability measured over a given frequency range may 
be associated with changes in power spectral amplitude and/or shape. 
For example, Belloni \& Hasinger (1990) show that the {\it fractional}
RMS variability of Cygnus~X-1 is well correlated with changes in the
position of the
low-frequency break in the power spectrum, but note that these variations
occur on long time-scales and show no relationship to X-ray flux. 
It seems unlikely that the strict linear relationship we have
found between RMS variability and flux could be caused by changes in
power-spectral shape over the entire 0.1--10~Hz frequency range, because
the power spectrum over this range is not a simple power law. 
Instead, the simplest explanation is that only the
power-spectral amplitude changes.  We can confirm this interpretation of
the $\sigma$-flux relation by plotting power
spectra which are binned according to lightcurve flux. 
Fig.~\ref{fig:fluxps} shows flux-binned power spectra of Cygnus~X-1, 
obtained by averaging the power spectra of lightcurve segments of 32~s 
duration,
according to whether the segment flux occupies the first or fourth quartile
of the overall distribution of fluxes. 
The resulting power spectra correspond to fluxes of 3092~count~s$^{-1}$
and 3679~count~s$^{-1}$ for the first and fourth flux quartiles
respectively, and show that the amplitude of the broadband noise variability
responds to flux in the same way across the entire power spectrum. 
There does not appear to be any systematic change in power spectral shape.
We obtain a similar result for SAX~J1808.4-3658. \\

\section{The flux-dependent variability of AGN}
Given the strong similarities between the power-spectral shapes of AGN
and XRBs, which suggest that the same variability processes are at work,
independent of the mass of the central object, we might expect AGN
to show the same flux-dependent X-ray variability as XRBs.  To test this
possibility, we obtained quasi-continuous lightcurves of $\sim$days duration
for three Seyfert galaxies,
NGC~4051, NGC~5506 and MCG-6-30-15, observed in Dec 1996, Jun 1997
and Aug 1997 respectively and available in the {\it RXTE} public
archive.  Because PCUs 3 and 4 were switched off for much of the duration
of each observation, we only use data from PCUs 0, 1 and 2. 
We make the lightcurves from Standard~2 data,
in the 2--10~keV energy band, using
data from only the top layer of each PCU in order to minimise the
contribution of instrumental background which adds to the Poisson noise
level, applying standard selection criteria (e.g. as described in Uttley
1999) to extract good time intervals, and estimating background
lightcurves using the L7 model. 
The observations are of total useful exposures
$\sim75$~ks, 90~ks and 340~ks for NGC~4051, NGC~5506 and MCG-6-30-15
respectively.  We made power spectra from continuous lightcurve segments
of length $\sim2500$~s (corresponding to periods between Earth occultations of
the source), and measured the Poisson-noise-subtracted integrated power in the
$5\times10^{-3}$--$5\times10^{-4}$~Hz frequency range. \\
Due to the faint
nature and lower variability (relative to XRBs) of AGN on the
time-scales sampled, we average the integrated power into two flux bins,
corresponding to mean segment fluxes lying below or above the mean flux of
the entire observation.  The resulting integrated powers (including standard
errors) are shown in Table~\ref{tab:agnrms}.
Note that we do not directly determine the RMS variability, $\sigma$ for each segment,
because statistical fluctuations in the true Poisson
noise level, combined with the relatively small amount of high-frequency
power seen in AGN lightcurves, sometimes lead to negative integrated powers after
subtraction of the noise estimate.  Segments with negative integrated
power cannot be used to determine $\sigma$, but can be used to determine
the mean $\sigma^{2}$ and its error (note that the fractional RMS
can be estimated from the mean $\sigma^{2}$ and is also shown in
Table~\ref{tab:agnrms}). \\
\begin{table*}
\caption{Flux-related changes in the variance of NGC~4051, NGC~5506 and
MCG-6-30-15.} \label{tab:agnrms}
\begin{tabular}{lccccccccc}
 & & \multicolumn{4}{c}{Low flux} & \multicolumn{4}{c}{High flux} \\
 & $\mu_{\rm tot}$ & $\mu$ & $n$ & $\overline{\sigma ^{2}}$ &
$\sigma_{\rm frac}
$ & $\mu$
& $n$ & $\overline{\sigma ^{2}}$ & $\sigma_{\rm frac}$ \\
NGC~4051 & 3.4 & 1.9 & 17 & $0.047\pm0.024$ & 11.4~\% & 5.3 & 16 &
$0.32\pm0.07$ & 10.7~\% \\
NGC~5506 & 28.1 & 25.5 & 20 & $0.21\pm0.05$ & 1.8~\% & 31.2 & 19 &
$0.41\pm0.11$ & 2.1~\% \\
MCG-6-30-15 & 12.2 & 10.3 & 73 & $0.28\pm0.05$ & 5.1~\% & 14.1 & 70 &
$0.56\pm0.07$ & 5.3~\% \\
\end{tabular}

\medskip
Mean data are shown for low and high flux segments
(corresponding to
segment fluxes below and above the mean for the entire observation,
given by $\mu_{\rm tot}$). 
$\mu$ is the mean flux (2--10~keV, count~s$^{-1}$) for each flux bin,
$n$ is the number of
segments in the flux bin, $\overline{\sigma ^{2}}$ is the mean variance
(count$^{2}$~s$^{-2}$) and $\sigma_{\rm frac}$ is the fractional RMS
variability ($\sigma_{\rm frac}=(\sigma^{2}/\mu^2)^{\frac{1}{2}}$)
\end{table*}
These data clearly show that the variance of the X-ray lightcurves of
all three AGN is dependent on X-ray flux, and furthermore the
relationship between the RMS variability and the mean segment flux seems
to be linear,
such that the fractional RMS variability remains approximately constant,
despite significant flux changes (e.g. $\sim$factor 3 for NGC~4051). 
There is not sufficient data to confirm the existence
of a constant flux component to the AGN lightcurves,
of the same relative strength as that seen in the XRBs, although we note that
the existence of such a component is not ruled out.

\section{Discussion}
\begin{figure}
\begin{center}
{\epsfxsize 0.9\hsize
 \leavevmode
 \epsffile{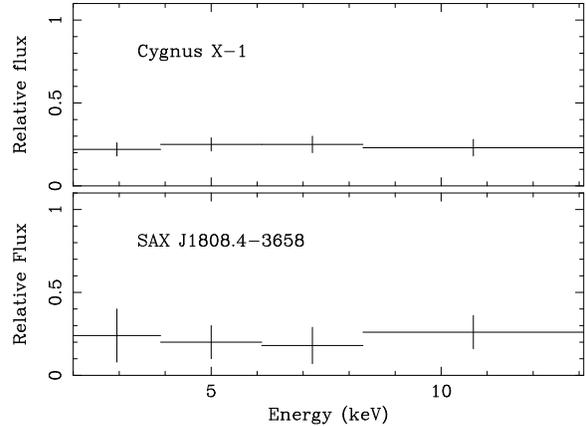}
}\caption{Energy-dependent flux of possible constant components to the 
lightcurves of Cyg~X-1 and SAX~J1808.4-3658, relative to the mean
flux in each band.}
\label{fig:conspec}
\end{center}
\end{figure}

\begin{figure}
\begin{center}
{\epsfxsize 0.9\hsize
 \leavevmode
 \epsffile{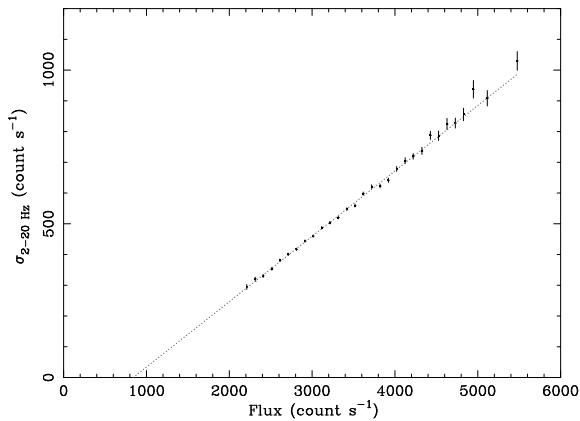}
}\caption{Flux-dependence of mean $\sigma$ (2-20~Hz band) for
Cyg~X-1, determined from
1~s lightcurve segments.} \label{fig:cyg1s}
\end{center}
\end{figure}

The relation between flux and the amplitude of broadband noise
variability which we have presented here, suggests that the lightcurves
of XRBs (and possibly AGN) are made from at least two components.  One
component shows a striking linear dependence of RMS variability on
flux, while the other component may contribute a constant RMS to the
lightcurve or, more simply, may not vary at all.  We can crudely estimate the
spectral shape of this second, possibly constant component by
determining the RMS-flux relation for lightcurves in four energy bands and
measuring the value of the intercept on the flux axis ($C$) for each band. 
In Fig.~\ref{fig:conspec} we plot the value of $C$ in each energy band
expressed as a fraction of the mean flux in that band.  The errors on
the relative flux are determined from errors in $C$ estimated from
fitting a linear model to the RMS vs. flux data for that energy band. 
Note however that because the fluxes in each energy band are strongly
correlated with one another (e.g. Maccarone, Coppi \& Poutanen 2000),
the {\it relative} uncertainty between energy
bands is likely to be significantly smaller than indicated by the error bars
we determine here.  With this caveat in mind, Fig.~\ref{fig:conspec}
suggests that, if $C$ represents the flux of a constant component to the
lightcurve, its spectral shape is very similar to that of the total
spectrum and hence the variable component.  \\
This result seems strange if we naively expect that any
constant components to the lightcurve should have a different origin and
therefore a different spectrum to the variable component. 
We note however, that provided that the temperature of the corona is
maintained, it will continue to Comptonise seed photons regardless of
whether it is dynamically variable (e.g. flaring) or `quiet' and non-variable.
Therefore the constant component to
the X-ray lightcurve may be associated with quiet regions
of the corona or patches of the corona above inactive regions of the
accretion disk (if variability is driven by variations in seed photon
number).  The $\sim25$\% of total flux which the constant component
contributes could indicate a $\sim25$\% covering fraction of quiet
coronal regions (this may represent a time-averaged covering fraction, since
coronal regions might switch from being quiet to being variable).  An
alternative possibility is that the constant component is due to Thomson
scattering of the primary continuum photons in extended,
hot gas with radius $R>10$~light-seconds to smear out any variability,
temperature $T\sim10$~keV so that no absorption edges are observed
and optical depth $\tau\sim0.25$.  However, a plasma of this extent
should be a strong source of thermal bremsstrahlung
emission.  Assuming a spherical geometry and uniform density,
the bremsstrahlung luminosity
from the extended plasma region would scale linearly with its radius so that
such a region would contribute $>10$\% of the X-ray
luminosity of the system in bremsstrahlung emission
and is ruled out by observations. \\
The variable component of the lightcurve shows a linear dependence of
RMS variability on flux, which is not consistent with the expectations
of simple shot-noise models, which predict that the power spectrum is
stationary.  At first glance, the obvious explanation is that
the parameters of the shot-noise process are changing on short
time-scales.  For example, if the average amplitude of shots varies on 
short time-scales, the flux and RMS variability will vary in proportion
to the shot amplitude.  Hence the linear relation between RMS
variability and flux reflects the linear relationship of both variables
to a single underlying varying parameter, the shot amplitude.  However,
this model cannot simply explain why the RMS variability is dependent on
flux on all measured time-scales, a fact which becomes apparent
if we plot the RMS-flux relation for Cygnus~X-1 using 1~s
time bins, determining $\sigma$ in the 2--20~Hz frequency range (see
Fig.~\ref{fig:cyg1s}).  The fact that the linear relationship between
$\sigma$ and flux extends to a greater range of fluxes than are obtained
by determining the relation using 10~s segments, implies that the
linear RMS-flux relation applies to flux variations on
time-scales as low as 1~s (which are averaged out by the 10~s segments
used earlier).  Therefore, if shot parameter variations are the cause of
the RMS-flux relation, they must occur on as broad a range of
time-scales as the variations in the lightcurve itself (at least down to
1~s), and we are left with the circular problem of trying to explain
the broadband noise in X-ray lightcurves with a model which itself requires
a broadband noise component (to describe the variation in shot parameters)
in order to be consistent with the data. \\
The simplest solution may be to discard the standard shot-noise models 
altogether.  Instead, the linear dependence of RMS variability on flux may be
more simply explained if the lightcurves are
instead made `from the top down', with the primary cause of variability
being large, long-time-scale variations on which the shorter-time-scale
variations are later superimposed.  For example, large-scale energy releases in
the corona (e.g. through magnetic reconnection) might further sub-divide
into a fractal structure, where the energy emitted by each
sub-unit is proportional to the energy of its parent unit, but the
time-scales for energy emission and the number of sub-units remain
independent of total energy content (perhaps related to characteristic
time-scales in the corona or disk). \\
Interestingly, a linear dependence of RMS variability on flux is a natural
outcome of the mechanism for producing red-noise variability proposed by
Lyubarskii (1997), where fluctuations in mass accretion rate at
different disk radii are propagated through the disk to produce
variations on all time-scales in the inner disk (and associated corona).
 In this model, fractional variations in mass accretion rate are produced
on time-scales greater than the viscous time-scale, so that longer time-scale
variations are produced in the outer disk and shorter time-scale
variations in accretion rate are superimposed on them as they propagate
inwards.  Thus, a linear RMS-flux relation will be produced if the 
fractional amplitude of accretion rate variations
is independent of the actual accretion rate (as suggested by the model,
which assumes that changes in accretion rate are caused by fractional
fluctuations in the disk viscosity parameter).  Either of these
suggested models might, in principle, explain the RMS-flux relation we
see, but we leave a detailed comparison of their predictions with
observations for a future work.
 
\section{Conclusions}
We have shown that the amplitude of the broadband noise variability
in the lightcurves of the black hole candidate Cyg~X-1 and the
millisecond X-ray pulsar SAX~J1808.4-3658 is dependent on flux,
in that the RMS variability for a given segment of the lightcurve scales
(on average) linearly with the segment mean flux.  The linear relation
between flux and RMS variability has a positive offset on the flux axis,
suggesting the existence of a second, constant-flux component to the
lightcurve which contributes $\sim25$\% of the total flux.  The shape of the
power spectrum (at least on short time-scales) remains independent of
flux.  The X-ray lightcurves of AGN are also consistent with the same linear
scaling of RMS variability with flux, suggesting that this behaviour is
an intrinsic feature of the broadband noise, which is characteristic of the
lightcurves of compact accreting systems across at least 
6 decades of central object mass. \\
The spectrum of the constant component to the lightcurve is similar to
the total spectrum, suggesting that it may correspond to quiet,
non-variable regions
in the X-ray emitting corona.  The unusual behaviour
of the variable component of the lightcurve,
is inherently difficult to explain using standard shot-noise
models.  An alternative, `top down' approach is more suitable, where the 
longest-time-scale variations precede the smallest. Possibilities
include large coronal flaring regions which break down into a fractal
structure of smaller sub-units, or models where variability is caused by
variations in mass accretion rate which propagate through the disk, so
that shorter time-scale variations from the inner parts of the disk are
superimposed on longer time-scale variations from further out. \\
If the linear RMS-flux relation is indeed common to the broadband noise
variability of all
compact accreting systems, it should provide a useful tool for
separating out components of lightcurves which are not related to the
broadband noise component.  It will also prove interesting to
see how other features of XRB power spectra (such as QPOs) scale with flux.

\subsection*{Acknowledgments}
We wish to thank the anonymous referee for useful comments. 
This research has made use of data obtained from the High Energy
Astrophysics Science Archive Research Center (HEASARC), provided by
NASA's Goddard Space Flight Center.

\bsp
\end{document}